\documentclass[conference]{IEEEtran}
\IEEEoverridecommandlockouts 
\usepackage{cite}\ifCLASSINFOpdf
   \usepackage[pdftex]{graphicx}
   \usepackage[pdftex]{hyperref}
   \graphicspath{{./Images/}}
   \DeclareGraphicsExtensions{.pdf,.jpeg,.png}
\else
\fi
\usepackage[cmex10]{amsmath}

\usepackage{multirow}
\usepackage{booktabs,colortbl}

\begin{document}
%
\title{A predictive pan-European economic and production dispatch model for the energy transition in the electricity sector}

\author{%

\IEEEauthorblockN{Laurent Pagnier}
\IEEEauthorblockA{Institute of Theoretical Physics\\
\'Ecole Polytechnique F\'ed\'erale de Lausanne\\
Lausanne, Switzerland\\
laurent.pagnier@epfl.ch}
\and
\IEEEauthorblockN{Philippe Jacquod}
\IEEEauthorblockA{School of Engineering\\
University of Applied Sciences of Western Switzerland \\
Sion, Switzerland\\
philippe.jacquod@hevs.ch}

%


}


\maketitle

\begin{abstract}
The energy transition is well underway in most European countries. It has a growing impact on electric power systems as it dramatically modifies the way electricity
is produced. To ensure a safe and smooth transition towards a pan-European electricity production dominated by renewable sources, it is of paramount importance 
to anticipate how production dispatches will evolve, to understand how increased fluctuations in power generations
can be absorbed at the pan-European level and to evaluate where the resulting changes in power flows will require significant grid upgrades. To address these issues,
we construct an aggregated model of the pan-European transmission network which we couple to an optimized, few-parameter 
dispatch algorithm to obtain time- and geographically-resolved 
production profiles. We demonstrate the validity of our dispatch algorithm by reproducing historical production time series for all power productions in fifteen 
different European countries. Having calibrated our model in this way, we investigate future production profiles at later stages of the energy transition -- determined by 
planned future production capacities -- and the resulting interregional power flows. We find that large power fluctuations from increasing penetrations of renewable
sources can be absorbed at the pan-European level via significantly increased electricity exchanges between different countries. We identify where these increased
exchanges will require additional power transfer capacities. We finally introduce a physically-based economic indicator which allows to predict future 
financial conditions in the electricity market. We anticipate new economic opportunities for dam hydroelectricity and pumped-storage plants.
\end{abstract}
%
\begin{IEEEkeywords}
Electricity market, power generation dispatch, power Transmission.
\end{IEEEkeywords}

\section{Introduction}

Most European countries are now engaged in the energy transition whose ultimate goal is to meet energy demand from human activities solely with renewable energy
sources (RES). In its current intermediate stages, the transition steadily increases the penetration of nondispatchable electricity productions, which results in large uncontrolled fluctuations in power generation. The development of RES in Europe follows from strong public investments and incentives which temporarily bias the electricity market. 
RES have negligible marginal cost and consequently their increasing penetration artificially lowers electricity prices below production costs for many other power generations.
Simultaneously, new RES such as solar photovoltaics and wind turbines have undispatchable, strongly fluctuating productions which need to be counterbalanced by
controllable, dispatchable productions and electrical energy storage solutions. In the context of the energy transition it is therefore of key importance to understand how
RES penetrations can be increased without jeopardizing the dispatchable productions required by the next level of RES penetration. 

One of our main interests here is the hydroelectric sector, with its fast dispatchable and fully controllable dam productions as well as pumped-storage (PS), the only sizeable,
mature storage solution to date. Hydroelectricity seems like an ideal partner to RES in the context of the energy transition 
and one may anticipate that further investments in new PS or
higher power dam facilities would significantly help to absorb increased production fluctuations from larger RES penetrations. However, somewhat ironically, the current 
low electricity prices penalize investments in new hydroelectric facilities -- with today's economic conditions in the European electric sector, RES jeopardize the future of 
hydroelectricity, arguably one of its main and most reliable future partner. To evaluate scenarios for the energy transition it is therefore of paramount importance to 
evaluate whether this trend will continue, and if yes, for how long, and determine if and when the precious flexibility of hydroelectric production will be again rewarded.
To achieve that, one needs a reliable dispatch model for all types of electric productions as well as a reliable economic indicator.
The current electricity market requires tools for financial analysis with increased precision
to identify the need for further production investments and the returns they will generate. Of particular interest is to try and implement transparent dispatch models
at the level of the pan-European grid that rely as weakly as possible on highly volatile political, economical or financial predictions. 

In this manuscript, we develop an integrated physico-economical power dispatch model relying on physical constraints for electric power productions,
on published future production capacity developments and on basic, demand-supply economical laws only. We demonstrate the validity of our model by reproducing
rather accurately historical 2015 electricity production profiles for all power production types in nineteen European countries. We argue that our model will become
more and more accurate as the energy transition progresses and investigate European dispatches as well as intercountry exchanges for 2030. This allows us to
identify the needs for increased grid capacity, for further storage capacity as well as future rules of engagement for hydroelectric dam power plants. Additionally,
we clarify the financial conditions prevailing in the electricity market in the forthcoming stages of the energy transition.

The main limitation of our approach is related to uncertainties in future installed production capacities in European countries, which will depend on more general economic and 
financial conditions in Europe and in the world as well as on future political and societal decisions. However, regardless of these mostly unpredictable conditions, we argue that 
our dispatch and revenue evaluation model retains its validity, provided production capacities are adapted to their true evolution. In other words, our model 
qualitatively predicts production dispatches and revenues for given production capacities. Accordingly, feasibility studies should consider various scenarios for future production
capacities to investigate which one presents the best operational and financial perspectives. 

The manuscript is organized as follows. In Section II we discuss our aggregated pan-European power grid model, the optimal power flow and the parameters 
on which our dispatch model is based.
In Section III we calibrate these parameters by reproducing historical data for the year 2015. In Section IV,
we apply our model to one ENTSO-E scenario for future European
electric power production capacities for the year 2030. Our results show how electric power is transferred across the continent as different meteorological conditions prevail, 
and from this, we infer the magnitude of intercountry power flows. In Section V we introduce the residual load as an economic indicator which allows us to evaluate 
normalized future revenues. As an example we calculate future revenues for pumped-storage hydroelectric power plants in Germany.
Conclusions and discussions of our results and our model are presented in Section VI.

\section{An aggregated model for future pan-European electricity dispatch}

We develop an equivalent model to determine future power dispatches in the pan-European power grid at different stages of the energy transition. Equivalent aggregated models have a relatively long history \cite{ward1949equivalent,tinney1987adaptive,papaemmanouil2011reduction}. They are standardly used for systemic investigations
such as ours, where precise details of power flows are not crucial (as opposed to, say, grid stability investigations) and exact, geographically resolved production and consumption
data are hard to obtain.

\subsection{An aggregated pan-European electric grid}

\begin{figure}
\center
\includegraphics[width=0.9 \columnwidth]{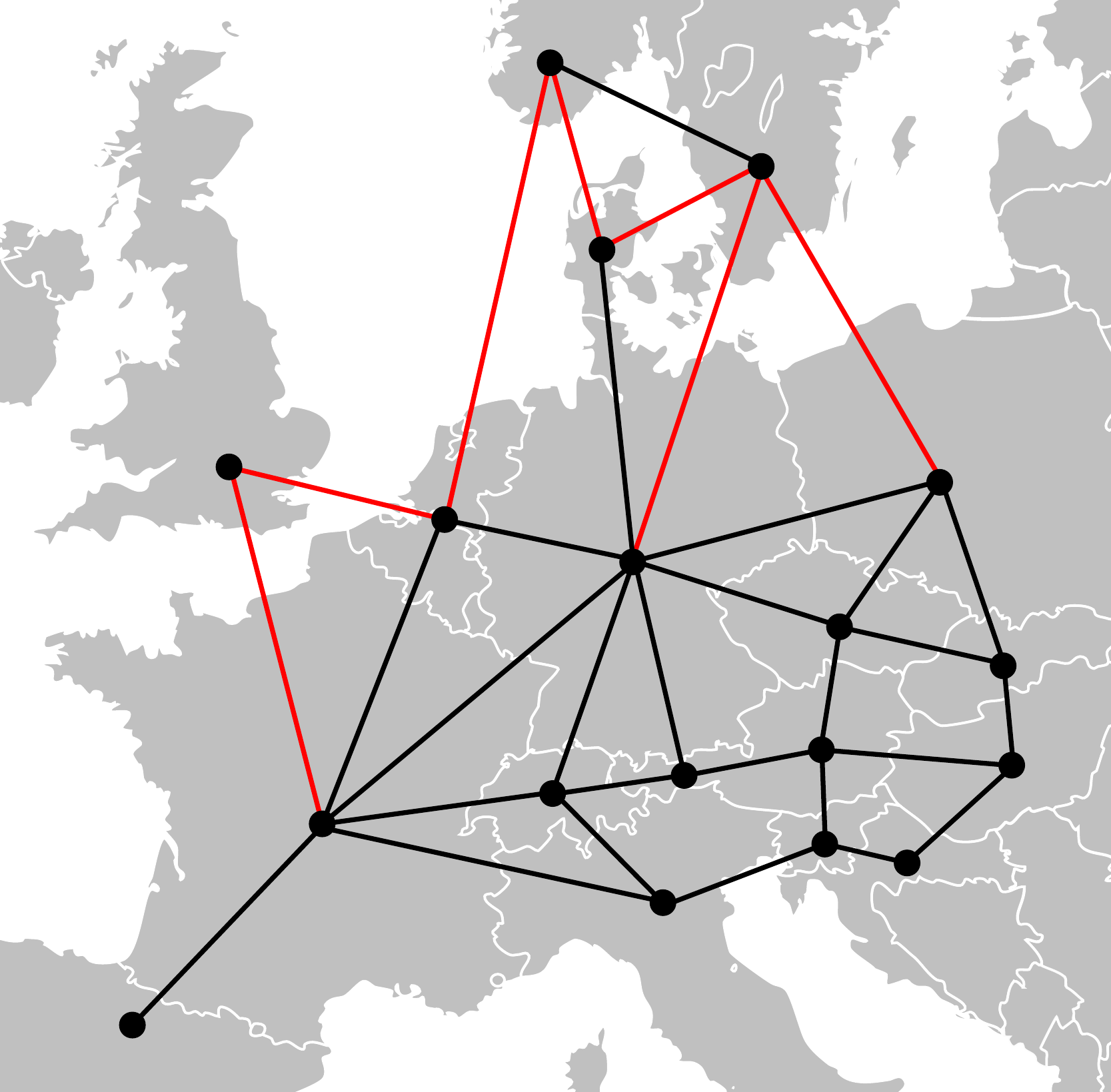}
\caption{Aggregated model of the Central and Northern European grid. Each node represents a dispatch zone. The lines represent interconnections: AC connections are in black and DC connections in red.}\label{euro_grid}
\end{figure}

Fig.~\ref{euro_grid} shows our aggregated European grid, with each node representing an independent dispatch zone (Portuguese consumption and production are 
included in the Spain node). Aggregated lines have admittances obtained via a standard reduction method\cite{shi2015novel} and thermal limits given by the sum of the physical lines they represent. The power flows are computed in the DC lossless approximation \cite{gomez2008electric}. 

\subsection{Productions and Consumptions}

Consumptions and productions are aggregated within each dispatch region and attributed to the corresponding node. Power productions are subdivided into two sets. They are,
\begin{itemize}
\item{Non-flexible productions, mostly consisting of run-of-the-river (RoR), solar photovoltaics (PV) and wind turbine productions. The remaining non-flexible productions are grouped into "miscellaneous productions". Note that RoR is in principle flexible, at least to some extent, 
however we neglect curtailment and consider that, as for PV and wind turbines, RoR production is
determined by weather/seasonal conditions only.}
\item{Flexible productions: We classify them into 6 types, which are (i) dam hydroelectricity, (ii) pumped-storage hydroelectricity (which can be positive as well as negative, but always
counted as a production), (iii) gas and oil, (iv) nuclear, (v) hard coal and (vi) lignite productions.}
\end{itemize}

\begin{figure}[h]
\center
\includegraphics[width=0.5\textwidth]{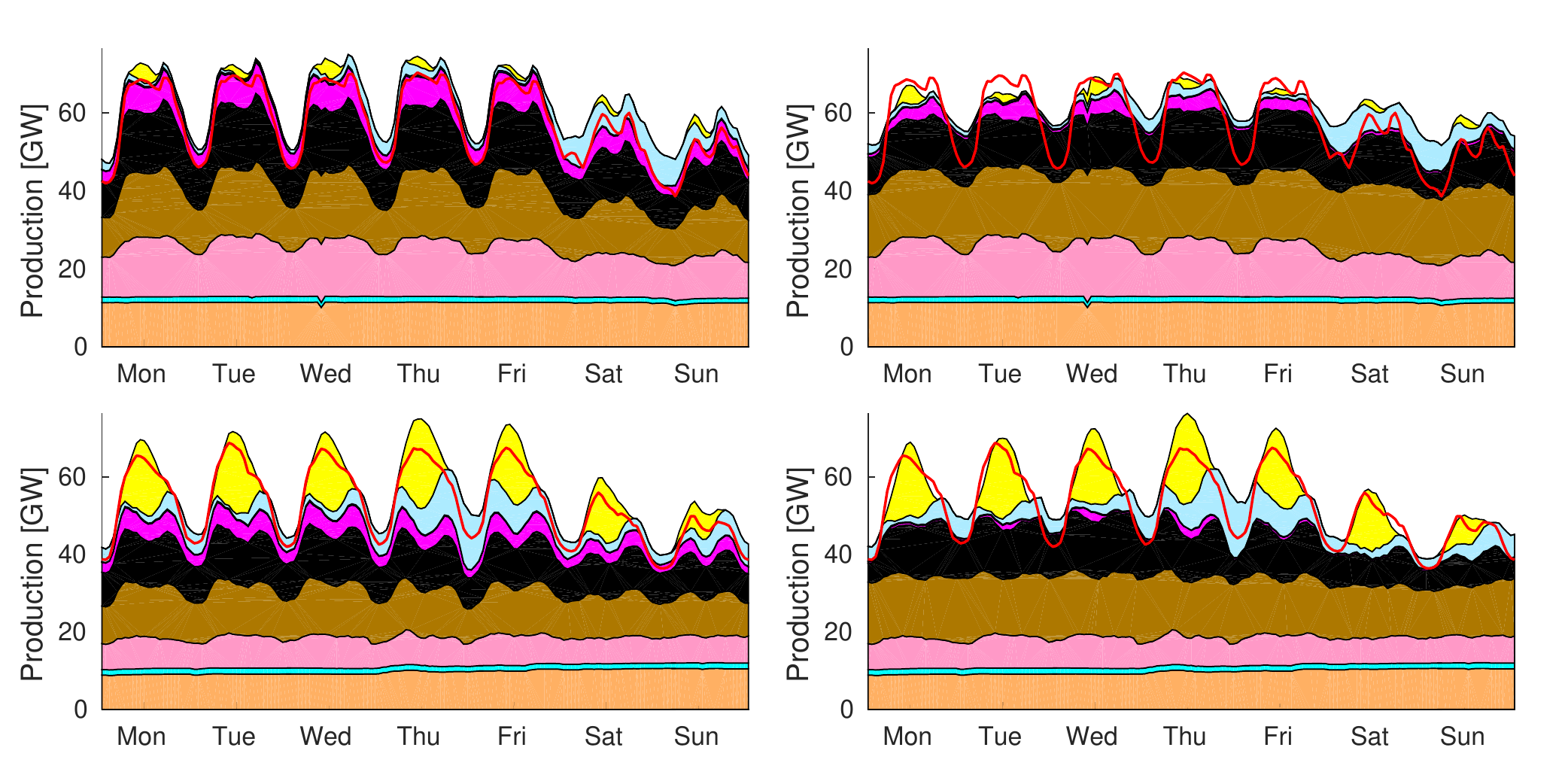}
\includegraphics[width=0.5\textwidth]{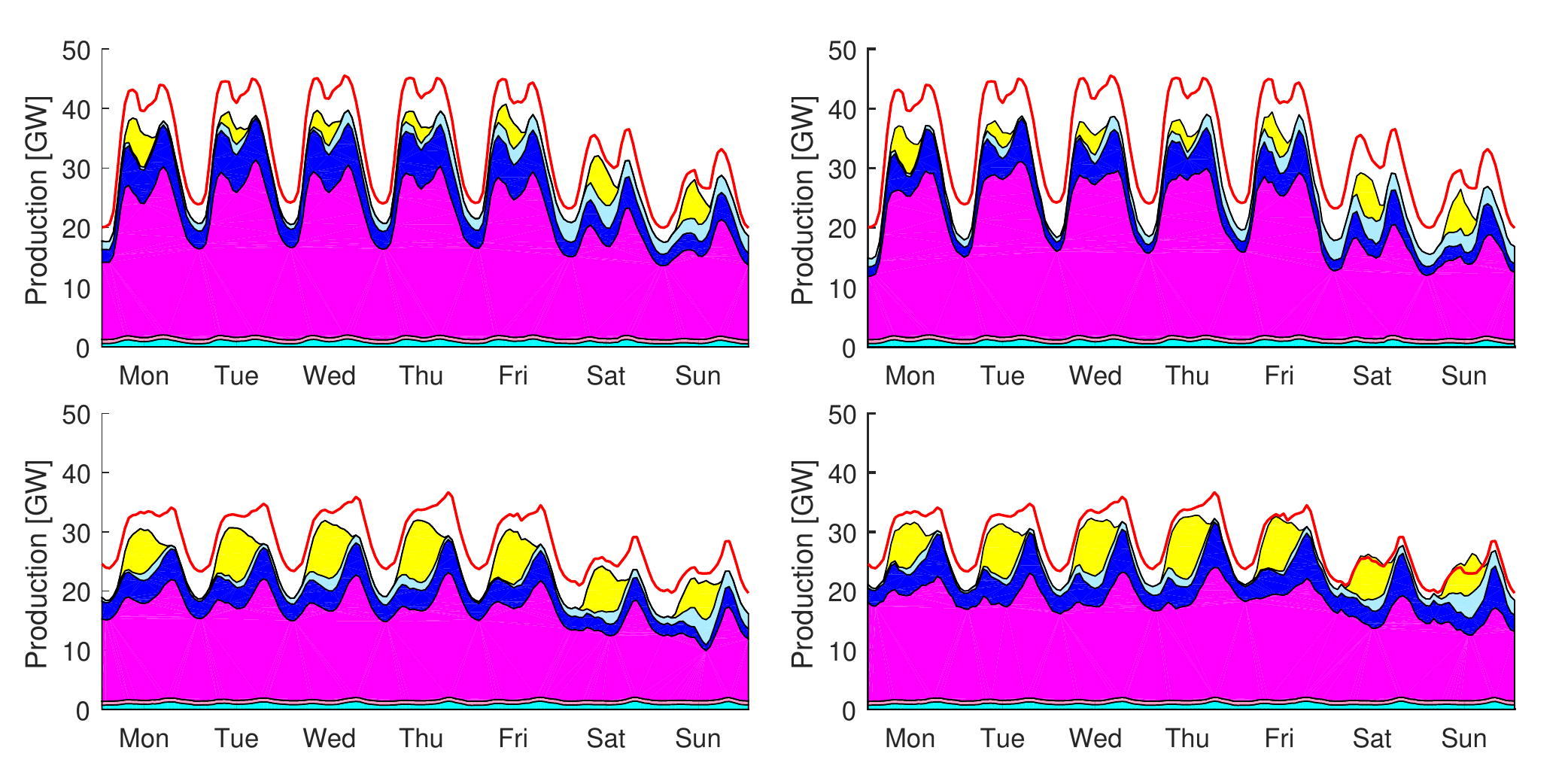}
\caption{Dispatched (left) and actual 2015 (right) production of Germany (top two rows) and Italy (bottom two rows) 
for a winter week (first and third rows) and a summer week (second and fourth rows).
Production types are: nuclear (orange), RoR (cyan), miscellaneous (pink), lignite (brown), hard coal (black), gas (purple; including hard coal in Italy), 
dam (blue), wind (light blue) and PV (yellow). The red curve indicates the 2015 national consumptions.}\label{DE_dispatch}
\end{figure}

 For each zone and at each time, we define the residual loads $R_i(t)$ as the difference between the consumption and the non-flexible productions, 
\begin{equation}\label{eq:rl}
R_i(t) = L_{i}(t)-P_i^{\mathrm{inflex}}(t),
\end{equation}
where $L_i(t)$ and $P_i^{\mathrm{inflex}}(t)$ respectively give the load and the sum of the non-flexible productions at time $t$ in the $i^{th}$ zone.
In our approach, non-flexible sources produce according to weather and seasonal conditions, and only flexible productions are dispatchable. 
Our task is therefore to dispatch all flexible productions so that their production is equal to the total residual load at all times - this is equivalent to satisfy the 
balance condition that consumption is equal to production at all times. 

The association of European transmission grid operators (ENTSO-E) provides data on historical production and load profiles and installed capacities in the different countries \cite{entsoe2015platform} and forecasts for annual RES productions~\cite{entsoe2015scenario,entsoe2015tyndp} that we use to set up our model.

\subsection{Economic dispatch}

A large number of different optimized power flows exist \cite{czisch2005szenarien,schaber2012transmission,comaty2014ist,rodriguez2015cost,schwippe2013pan}. Our dispatch algorithm follows a merit order. The latter is based, first, on marginal costs, $a^k$, specific to each production type, $k$. Second, we introduce effective parameters
in the form of repulsion costs, $b^k$, which progressively increase the total production cost as the production increases and reaches its maximal possible value. Such repulsion
costs do not directly correspond to any real economic cost, however we found that they are necessary to smoothen production curves and reproduce historical time series faithfully. 
With these two parameters for each of the 
six different flexible productions, our model has a total of 12 parameters that need to be calibrated.

\begin{figure}[!t]
\center
\includegraphics[width=0.5\textwidth]{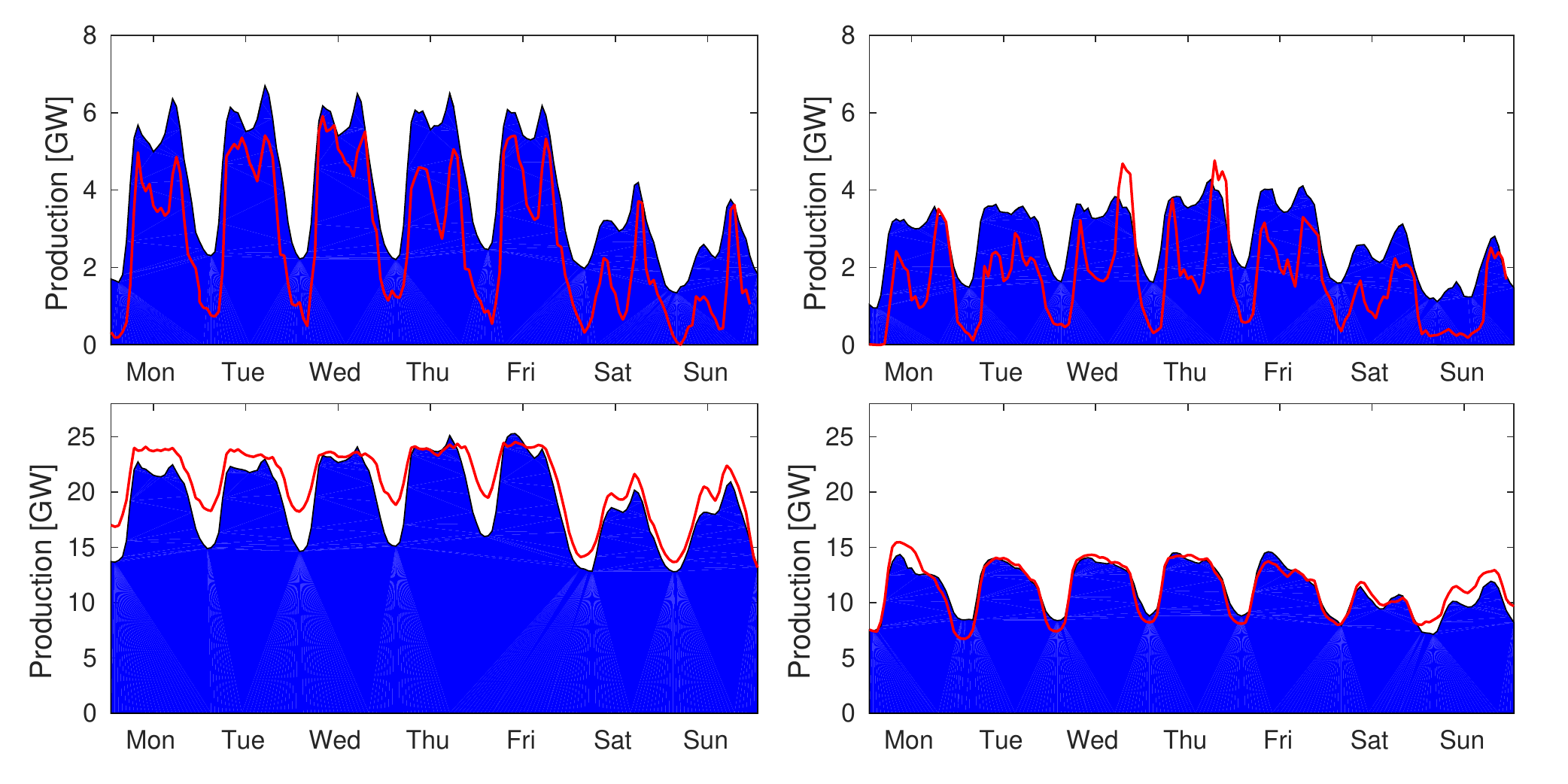}
\caption{Dam production of Switzerland (top) and Norway (bottom) for a week in winter (left) and summer (right) in 2015.
Dispatched productions are displayed in blue and actual 2015 production profiles are in red.}\label{hydro_dispatch}
\end{figure}

The production cost in the $i^{th}$ zone at each time 
step $\Delta t=1\mathrm{h}$ is given by a sum over the marginal and repulsion costs for all production types as
\begin{equation}
  W_i(t) = \sum_k \left[a^kP_i^k(t) + b^k\frac{P_i^k(t)^2}{P_{\max i}^k}\right]\Delta t,
\end{equation}
where $P_i^k(t)$ is the power generated by a given production type labelled $k$, in a geographical zone labelled $i$, at time $t$, and $P_{\max i}^k$ is the corresponding installed capacity. Our algorithm is based on an optimal power flow which 
determines the production profiles $\{P_i^k(t)\}$ minimizing the total, annual generation cost 
\begin{equation}\label{eq:gains}
W(\{P_i^k(t)\}) = \sum_{i,t} W_i(t),
\end{equation}
under the following technical constraints:

\paragraph{Power limits} $P_i^k(t) \le P_{\max i}^k$, $\forall t$; the power generated never exceeds its maximal installed capacity.

\paragraph{Ramp rates} $| \partial P_i^k(t)/\partial t| \le \Gamma_i^k$, $\forall t$; each production type has a maximal ramp rate $\Gamma_i^k$
at which the production increases or decreases. These ramp rates are similar, but not exactly equal, to the real, technical rates. We adapted them 
slightly when calibrating our model, to reproduce historical production time series better. 

\paragraph{Internodal power flows} $|P_{ij}(t)| \le P_{ij}^{\rm therm}$; they should never exceed the thermal limit $P_{ij}^{\rm therm}$ of the 
aggregated line between node $i$ and $j$ that carries them; when they do, a different dispatch must be implemented to 
correct this.

\paragraph{Dam storage}
Dam hydroelectric plants are constrained by the finiteness of their reservoir and the annual water intake into the latter.

\section{Model Calibration}\label{model_cal}

To calibrate the parameters in our model, we fixed non-flexible productions to those of 2015 and optimized the 12 parameters
in our model to reproduce true 2015 production data as faithfully as possible.
In Fig.~\ref{DE_dispatch} we show the result for a winter and a summer week in Germany and Italy, after the 12 parameters have been optimized. 
The agreement between dispatched and actual productions is excellent. We found comparable agreement between calculated and real 2015 productions
for all other countries in our aggregated model. Another level of complexity is brought about by dam hydroelectricity with its great flexibility. 
To illustrate that our dispatch model works even in that case, we show in Fig.~\ref{hydro_dispatch} 
the productions of Swiss and Norwegian dam hydroelectric plants during one week in summer and winter. Despite the inherent difficulty to dispatch dam hydro production, we see that our model captures most features of the 2015 production rather faithfully. Few discrepancies exist, in particular our calculation overuses flexibility in Norway in winter, 
which we attribute to mid- and long-term supply contracts whose effect cannot be captured by our model. Even with these few discrepancies,
we are unaware of another model that captures the national productions up to this level of detail at a European scale, including hydroelectric productions. 
From Fig.~\ref{DE_dispatch} and \ref{hydro_dispatch}, we conclude that our model is calibrated and fully valid. Its 12 free parameters having been fixed, we next 
use the model to investigate future scenarios of the energy transition. 

\begin{figure}[!]
\includegraphics[width=\columnwidth]{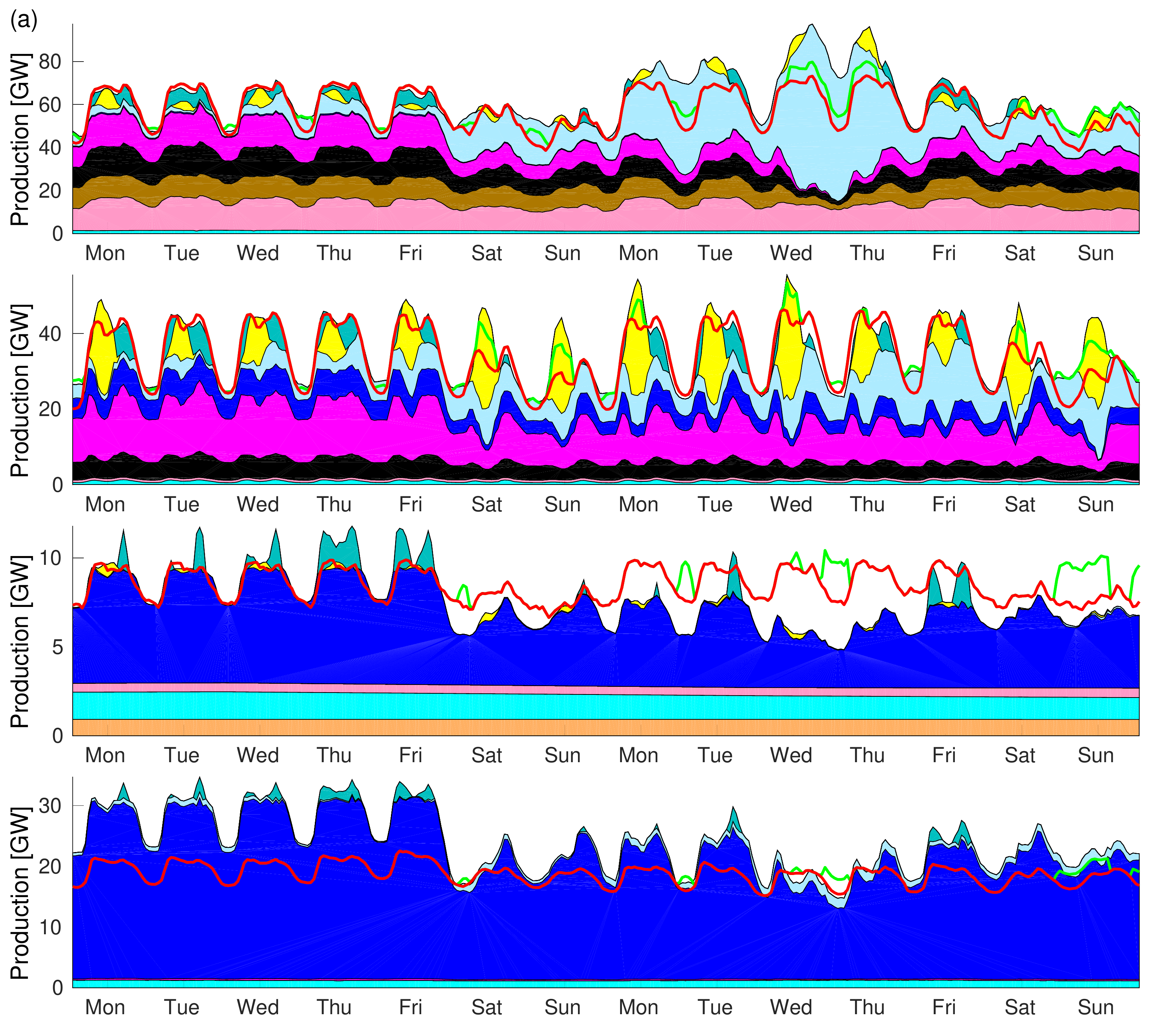}
\caption{Top to bottom: electricity productions of Germany, Italy, Switzerland and Norway for two winter weeks in 2030. Color convention is as in Fig.~\ref{DE_dispatch}, with additionally, pump-storage production (turquoise) and pump-storage consumption (pumping; green line).}\label{dispatch_30_W}
\end{figure}

\section{Future power dispatch}

Having calibrated our model, we next investigate how the flexible productions are dispatched, both geographically and in time, to handle future large penetration of RES in later stages of the energy transition. Our results are based on three assumptions. First, for production capacities in each country, we use the ENTSO-E scenario {\it 2030 Vision 4 of European Green Revolution} ~\cite{entsoe2015tyndp}. Second, non-flexible productions are obtained by rescaling their 2015 production profiles in direct proportion to their capacity evolution.
Third, we assume that 
consumption profiles will not be too different in 2030 from what they are now and use 2015 consumption profiles for each country in the aggregated model of Fig.~\ref{euro_grid}.
Obviously, our model can be used to check any other production and consumption scenario one may wish to implement. 

\begin{figure}[!]
\center
\includegraphics[width=1 \columnwidth]{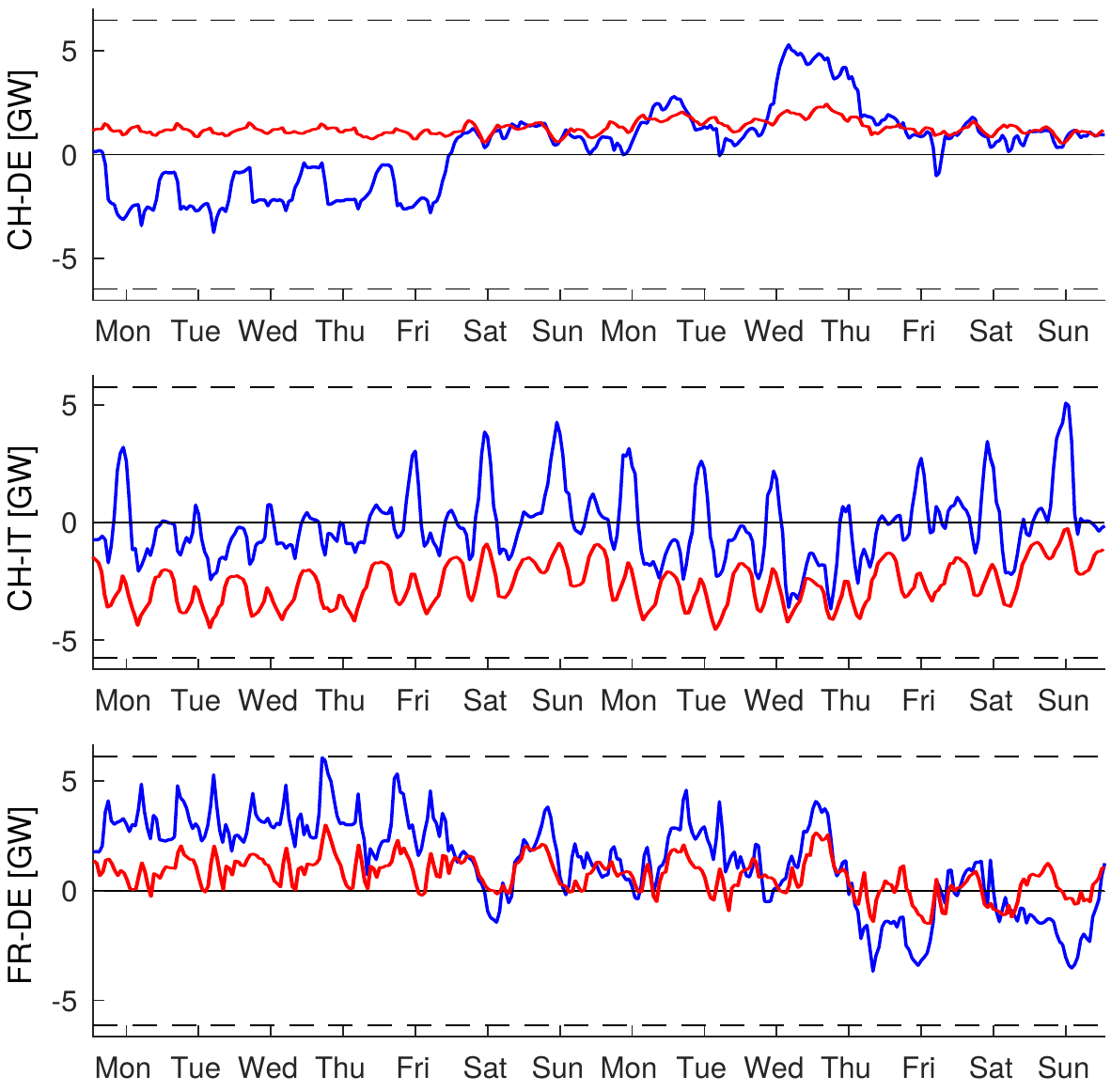}
\caption{Top to bottom: power flows between Switzerland and Germany, Switzerland and Italy and France and Germany in 2015 (red) and 2030 (blue). The thermal limit power of each connection is indicated by a dashed line.}\label{power_flows}
\end{figure}

Fig.~\ref{dispatch_30_W} shows the productions of Germany, Italy, Switzerland and Norway for two consecutive weeks in the winter of 2030. One sees first that when RES have low production (first five days), dam hydro productions are high to help supplying the demand for electricity. When RES productions are high, dam hydro production is significantly lowered. In particular, one sees that, with large RES productions, Switzerland continuously imports electricity during several consecutive days, which is never the case nowadays. 
Pump-storage hydro is additionally intensively used, as it produces a lot when RES produce little and consumes (pumps) when RES productions are high.
The yearly dam hydro production corresponds to the yearly water intake and as it is not expected to change significantly in the next two decades, the annual productions of Norway and Switzerland are comparable to the 2015 productions. Note that the total Swiss production diminishes a bit compared to 2015, 
which is due to the incomplete substitution of dismantled nuclear power by RES in the chosen ENTSO-E scenario for Switzerland.

Fig.~\ref{power_flows} shows the power flow of three important interconnects for the same two weeks. For comparison we added the power flows obtained for 2015 for the same period. We observe that more flexibility is asked of dispatchable productions. In particular, the flows in 2015 tend to have a dominant direction. For instance, the  CH-IT connection is used for Italian import only. In 2030, however, the increased Italian PV capacity results in a reversed flow across the  CH-IT interconnect.
It is clear that large RES productions induce increased power exchanges between European countries, often reversing the direction of the power flows and leading the 
latter regularly close to their thermal limits and sometimes in an unexpected direction.

\begin{figure}
\center
\includegraphics[width=0.9 \columnwidth]{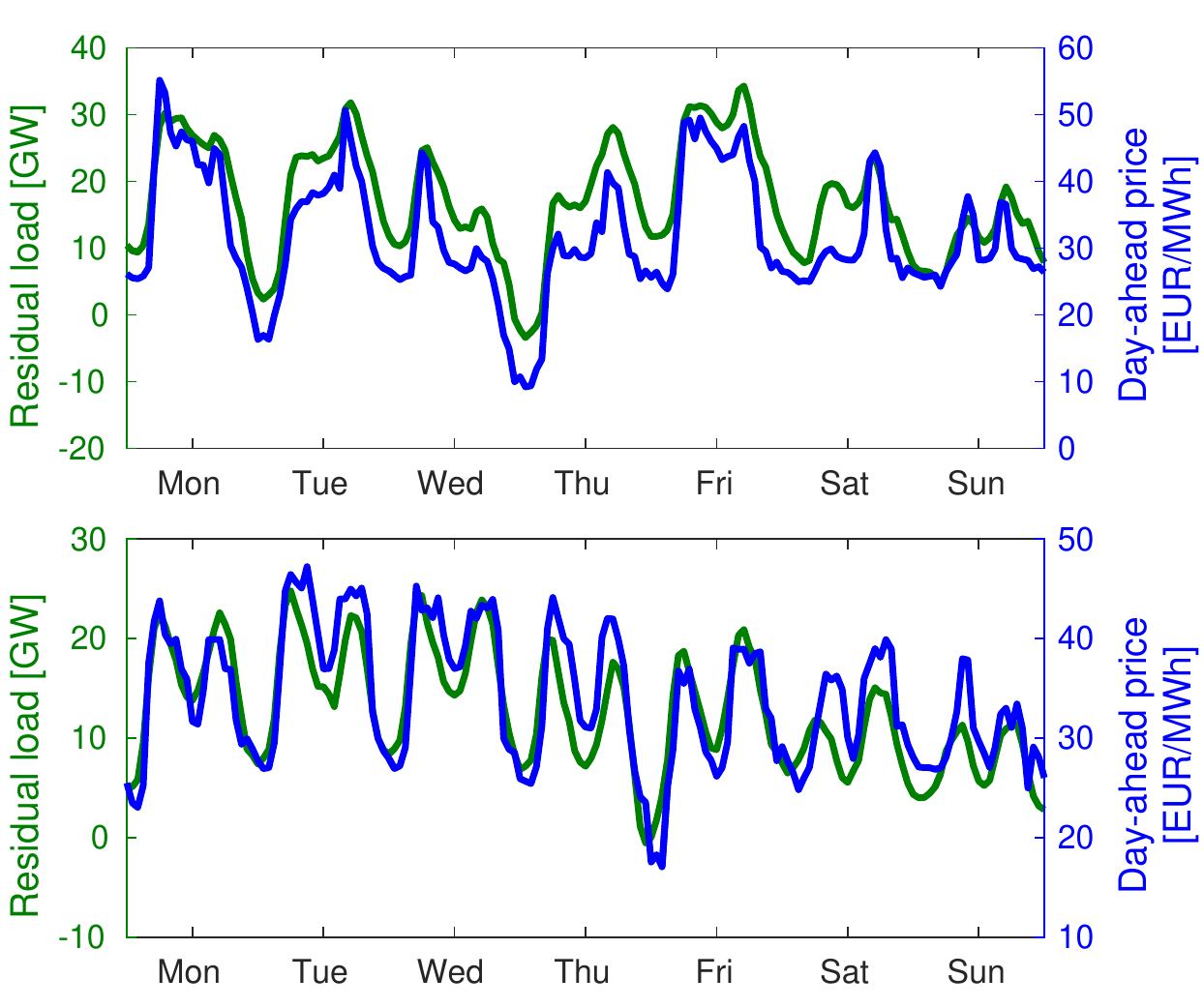}
\caption{German residual load (green line) and the German day-head electricity price (blue) for a winter (top) and summer (bottom) week in 2015. There exists a clear, almost perfect correlation
between the two quantities.}\label{rl_da}
\end{figure}

\section{Effective electricity price}
To anticipate changes and necessary upgrades to electric power systems in the light of the energy transition, a reliable economic indicator is needed which 
gives a qualitatively reliable estimate for the price of electricity. Here we deliberately 
choose to use  an indicator solely based on technico-physical conditions and not on highly speculative economic forecasts. As a matter of fact, such an economic indicator 
exists, which reflects quite clearly the law of supply and demand : it is the residual load $R_i(t)$ defined in Eq.~(\ref{eq:rl}). At a qualitative level, it is easily understood that when
$R_i(t)$ is low (high), the demand for flexible electricity and thus the price one is ready to pay for it are also low (high). More surprising is that the 
correlation between electricity prices and residual load is almost perfect quantitatively. This correlation was already observed in Ref. \cite{huber2010modeling} for the case of Germany.
Fig.~\ref{rl_da} shows that the residual load is strongly correlated with the day-ahead prices in Germany for two weeks, one in summer and one in winter. 
The correlation remains strong even when considering the whole year and we found a correlation coefficient between residual load and day-ahead electricity
price distributed as $r \in [0.5,0.9]$ for all countries in our aggregated model. Given that nowadays a major part of electricity transactions occur on the day-ahead market, 
and that in all likelihood, with the expiration and non-renewal of many long-term electricity contracts, this trend will be strengthen in the foreseeable future,
such large correlations suggests to introduce an effective electricity price based on the residual load as 
\begin{IEEEeqnarray}{rCl}\label{eq:price}
p_{\mathrm{eff},i}(t) = \alpha_i R_i(t) + \beta_i \, , 
\end{IEEEeqnarray}
with two parameters $\alpha_i$ and $\beta_i$ to be empirically determined from historical data. We obtained estimates
$\alpha_i \simeq$ 1 [EUR/(MWh $\cdot$ GW)] and
$\beta_i \simeq$ 20 [EUR/MWh] from recent historical data for Germany.

Having introduced this effective electricity price, it is now possible to investigate future economic conditions and opportunities with our model. 
To illustrate this, we evaluate future economic conditions for pumped-storage (PS) power plants. 
The revenue generated by a PS plant depends on its pump/turbine powers $P_{\mathrm{p}i}(t)$ and $P_{\mathrm{t}i}(t)$  
and the filling $S_{\mathrm{PS}_i}(t)$ of its reservoirs as
\begin{IEEEeqnarray}{rCl}\label{eq:ps0}
G & = & \sum_k p_{\mathrm{eff}}(t_k)[P_{\mathrm{t}_i}(t_k)-P_{\mathrm{p}_i}(t_k)]\Delta t \\
&\mathrm{s.t.} &\hspace{5pt} 0\le S_{\mathrm{PS}_i}(t_k) \le S_{\mathrm{PS}i}^{\max} \hspace{5pt}\, ,  \forall k \, . \label{eq:ps1}
\end{IEEEeqnarray}
At each time step $\Delta t=$1h, the reservoir filling evolves as
\begin{equation}\label{eq:ps2}
  S_{\mathrm{PS}_i}(t+\Delta t) = S_{\mathrm{PS}_i}(t) + [\eta P_{\mathrm{p}_i}(t)-\eta^{-1} P_{\mathrm{t}_i}(t)]\Delta t
\end{equation}
with a typical pump/turbine efficiency of $\eta = 0.9$. Including hydro pumped-storage defined by Eqs.~(\ref{eq:ps0}--\ref{eq:ps2}), our pan-European
aggregated model is similar to the 
power-node model of Ref.~\cite{Heu10}. 


\begin{figure}
\center
\includegraphics[width=1 \columnwidth]{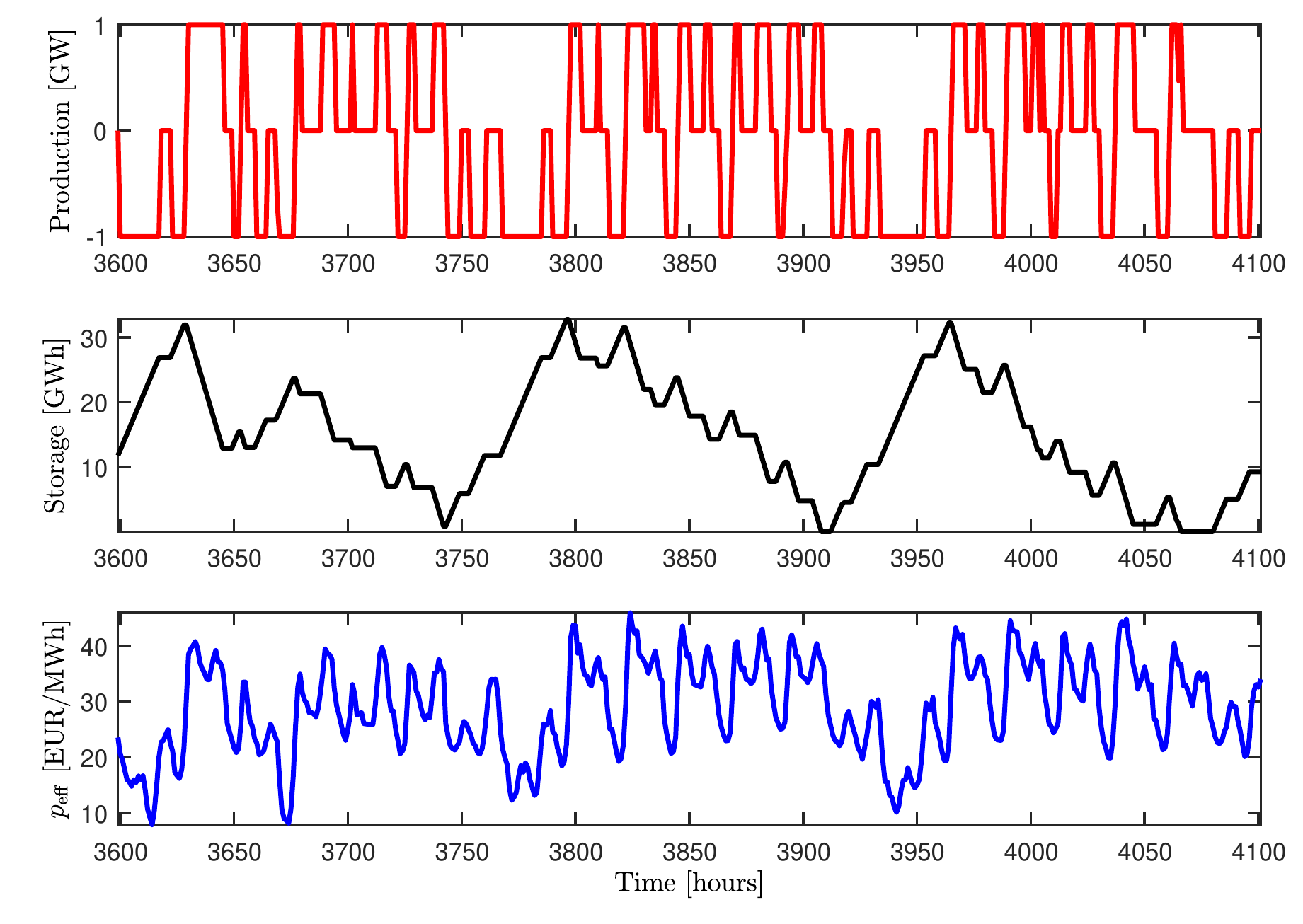}
\caption{Optimized production profile (top panel; $P>0$ correspond to production, $P<0$ to consumption/pumping), 
reservoir level (middle) and electricity price (bottom) for a 1 GW, 32 GWh PS plant.}\label{fig:ps}
\end{figure}

To obtain the PS production/consumption profile, we include its effective revenue, $G$ in Eq.~(\ref{eq:ps0}), in the total gain $W$ to optimize [see Eq.~(\ref{eq:gains})], 
and the constraints of Eqs.~(\ref{eq:ps1}) and (\ref{eq:ps2}) into our aggregated model. Fig.~\ref{fig:ps} shows time profiles for PS production/consumption, 
electricity prices and PS reservoir level for a fictitious 1 GW, 32 GWh PS plant. The production profile is, as expected, clearly correlated with the electricity price, and the constraints
on the reservoir level are met. 

\begin{figure}
\center
\includegraphics[width=1 \columnwidth]{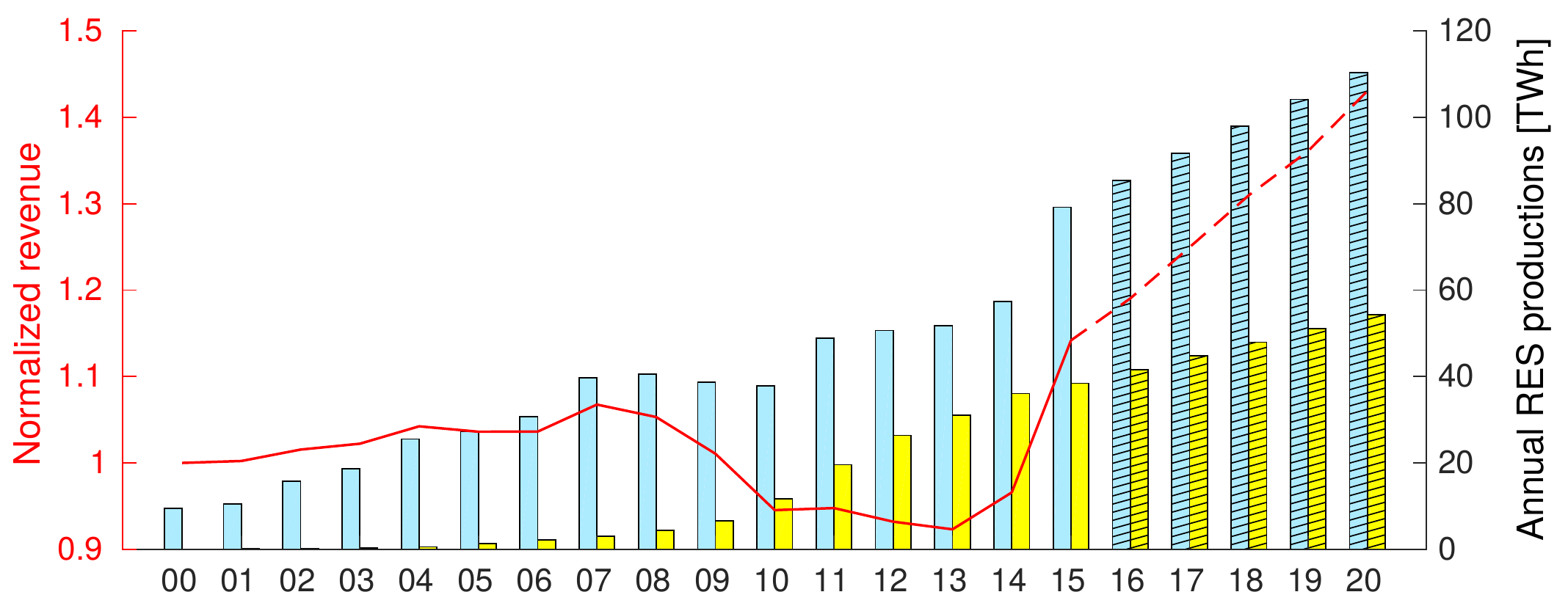}
\caption{Normalized revenue for PS (red line) superimposed on annual production for PV (yellow) and wind turbines (light blue)  in Germany.
A pump-storage efficiency of $\eta = 0.9$ each way is assumed.
}\label{fig:psgains}
\end{figure}

The PS revenue is further calculated using Eq.~(\ref{eq:ps0}) and we plot it in Fig.~\ref{fig:psgains} for the special case of Germany. Data are superimposed on
histograms depicting the annual RES electricity production for PV (yellow) and wind turbines (light blue).
We normalized the revenue with the 
revenue obtained from our dispatch model for the year 2000. We correctly obtain a significant revenue reduction from 2008 on, with a minimum around 2013, after which the
revenue increases again to get back to its pre-2008 value at around 2015-2016. The latter behavior is likely a bit premature, however, overall, our data 
qualitatively suggest that, (i) after a period of difficulties, PS will get back to its pre-2008 profit level rather soon, at least in Germany, and (ii) how soon PS gets back to 
larger profit margins depends mostly on how fast RES are substituted for fossil productions. 

\section{Conclusions}

We have constructed a pan-European model for the future electricity market. Using a mathematically well-defined merit order, we calibrated it so that it reproduces 2015 
production profiles. We investigated how productions will change up to 2030 and found that enhanced intercountry power exchanges will help absorbing 
large fluctuations of productions from PV and wind turbines. We introduced an effective electricity price and illustrated its predictive power by investigating revenues
generated by pump-storage facilities in Germany. Our results suggest that 
hydro pump-storage power plants will again generate comfortable profits in the future. How soon that will be depends 
mostly on the pace at which the energy transition proceeds.

\section*{Acknowledgment}

This work has been supported by the Swiss National Science Foundation.



\end{document}